# New categories of Safe Faults in a processor-based Embedded System

C. Gursoy[2], M. Jenihhin[2], A.S. Oyeniran[2], D. Piumatti[1], J. Raik[2], M. Sonza Reorda[1], R. Ubar[2]
1) Politecnico di Torino, Dip. Automatica e Informatica - Torino, Italy
2) Tallinn University of Technology - Tallinn, Estonia

*Abstract* - The identification of safe faults (i.e., faults which are guaranteed not to produce any failure) in an electronic system is a crucial step when analyzing its dependability and its test plan development. Unfortunately, safe fault identification is poorly supported by available EDA tools, and thus remains an open problem. The complexity growth of modern systems used in safety-critical applications further complicates their identification. In this article, we identify some classes of safe faults within an embedded system based on a pipelined processor. A new method for automating the safe fault identification is also proposed. The safe faults belonging to each class are identified resorting to Automatic Test Pattern Generation (ATPG) techniques. The proposed methodology is applied to a sample system built around the OpenRisc1200 open source processor.

1. **Introduction**

Electronic systems are increasingly adopted in safety-critical applications, where the effects of possible faults may cause significant damages, either from an economical point of view (e.g., in telecommunication systems) or in terms of consequences for the human users (e.g., in automotive systems). Hence, in these kinds of applications, it is strictly required to complement the usual design process with a dependability evaluation process, aiming at checking whether the system fulfills the dependability specifications. In the negative case, suitable counteractions have to be taken to improve the system dependability. Dependability evaluation [1] mainly corresponds to estimating the probability that the system produces a critical failure due to a fault. The dependability evaluation process is known to be complex, typically it starts by estimating the failure rate for every component of the system. When the component corresponds to a simple device (e.g., a resistor) the corresponding failure rate can be estimated resorting to available data bases (e.g., FIDES [15]). When it is a more complex device (e.g., a microcontroller, or a System on Chip (SoC), more sophisticated approaches have to be applied, taking into account the adopted technology as well as the mission profile (describing the kinds of stresses the component will be subject to). In any case, for each possible hardware fault affecting the component, we must associate two probability to it: the probability of occurrence the fault, and the conditional probability that, if the fault arises, it may cause a failure. By combining all these figures (Failure Modes and Effects Analysis, or FMEA) the dependability expert can basically estimate the dependability of the whole system. Since not all failures have the same severity, the analysis may also distinguish among different faults, basing on the severity of the failures they may produce (Failure Modes and Effects Criticality Analysis, or FMECA). For the purpose of this paper we ignore any distinction within the set of safe faults based on the criticality of the effects, which depends on the specific application.

Clearly, the system can include fault tolerance mechanisms (e.g., based on redundancy) and in-field test mechanisms (e.g., Software Test Library approach). The adopted solution should be able to guarantee that a sufficiently high percentage of possible faults (ideally, 100%) are detected. However, when computing this percentage (known as *Fault Coverage*) we should remove from the list of considered faults all those that for different reasons are known not to be able to produce any failure (*Safe Faults*). The set of Safe Faults for a given system includes the well-known *Untestable Faults*, that are usually caused by structural or sequential redundancy (which cannot be tested even by an exhaustive test) but it also contains faults that cannot produce any failure (and thus cannot be tested) due to the specific configuration of the system, which in some way limits the controllability and observability of each unit inside. It has been shown [4] that the percentage of Safe Faults may be significant (achieving 20% or 30% of the total number of faults in many cases), and it is thus crucial to be able to identify them. Similar figures have been obtained when analyzing the portion of a processor which is not used by a given application [5]. On the other side, systematic methods able to automate this step are rather immature, thus forcing most companies to perform it in a manual manner. The method described in [11] is potentially able not only to generate test programs for a processor, but also to identify safe faults given some constraints. Pruning the fault list used for assessing the fault coverage achieved by a functional test from safe faults may result in a significant computational time reduction. Moreover, any solution able to preliminarily identify Safe Fault will also be able to make the test generation step more effective. Following our previous work on the topic [2][3][4], in this paper we focus on the Safe Faults which may exist in an embedded system based on a pipelined processor (as it commonly happens in several safety-critical systems). Initially, an overview about the safe faults is shows (Section 2). After, we introduce some categories of Safe Faults (Section 3) that can be commonly found, and then highlight a method which is able to partly automate their identification, leveraging on existing commercial EDA tools (Section 4). Finally, we apply the proposed methodology on a sample system, and report experimental (Section 5) results to discuss the advantages and limitations of the proposed solutions (Section 6).

2. **Background**

The purpose of this Section is introduce the Safe Faults concept. The definition of Safe Fault is "*A fault which is guaranteed not to be able to produce any failure in the considered system*". Clearly, untestable faults (i.e., faults for which no test exists) are by definition safe faults. If a fault is a safe one, there is no reason for testing it. According to the above definition, different types of faults can be classified as safe. The *structurally (or combinationally) untestable faults*, are faults for which a test does not exist even if the combinational block where the fault is located is fully controllable and observable. An ATPG tool can identify these faults. The *sequentially untestable faults* are faults that do not belong to the previous group, but cannot be tested due to the sequential behavior of the circuit, for example, because the circuit cannot reach any of the states required for their test. Several works proposed techniques to automatically identify these faults [7][8][9][10][13][14][16][17]. *On-line functionally untestable faults* [2], are faults that do not belong to the previous groups, but cannot be tested in a functional manner in the operational conditions (i.e., without resorting to Design for Testability) the target device works in, as defined by the hardware configuration. Finally, the *application-dependent safe faults* are safe faults related to the software application run by the target system. The above classification is independent on the adopted fault model.

## 3. Categories of safe faults

In this section, some categories of safe faults in a processor-based embedded system are presented. All the described categories of safe faults are related to *constraints* that are valid in the target embedded system during the execution of a given application. In Section 4, we will show how these constraints can be used to identify the related Safe Faults.

**SPECIAL PURPOSE REGISTERS**. The first category of safe faults involves the usage of the *Special Purpose Registers (SPRs)* inside the processor. Special Purpose Registers can be used to configure the processor or to use the I/O peripherals. Each SPR is enabled by placing its identifier (ID) on a dedicated bus managed by an SPR addressing unit. The bus exiting the SPR addressing unit is able to address $2^n$ SPRs, where n is the parallelism of the bus. Usually the addressing capability is considerably higher than the number of SPRs really present in the processor. Therefore, there are numerous combinations of invalid IDs potentially generated by the SPR addressing unit but which will never to be generated. Moreover, it is possible that some of the SPRs are not accessed because they are not used by the specific application. Faults associated with the hardware that generates these never used SPR IDs will never be excited during the work of the processor, and thus belong to the category of safe faults. Since SPRs are located within the Register File, this module will never receive the IDs of the unused SPRs. Note, if a User Register is not used by the application, its ID is also never generated, and further safe faults may stem from this fact.

**DATA MEMORY ACCESS**. A second category of safe faults is associated with the *addressing of the memory* in the processor. The unit denoted as Memory Address Generator is able to address $2^n$ memory locations, where $n$ is the parallelism of the address bus. Typically, the memories used in embedded systems are smaller than the address space that can be generated by the *Memory Address Generator* unit. Therefore, there are numerous memory addresses which will never be generated during the embedded system life. This constraint will turn several faults in the *Memory Address Generator* unit into safe faults.

**PROGRAM COUNTER LOGIC**. The third category of safe faults is similar to the previous one but associated with the generation of addresses towards the code memory. In this case, a *Program Counter (PC) Generation unit* can exist within the CPU to generate addresses on $2^n$ bits, where *n* is the parallelism of the address bus to the code memory. The code memory used in these processors is normally smaller than the possible address space. Therefore, there are numerous invalid memory addresses. Once again, this turns into a constraint which leads to Safe Faults.

**INSTRUCTION DECODING LOGIC**. The fourth category of safe faults relates to the *decoding of the instruction opcode*. Assuming that the instruction opcode is represented on n bits, this represents an input to the decode unit. In theory, the processor can potentially decode $2^n$ possible instructions. Typically, the processor instruction set uses a lower number of instructions. Therefore, there are various invalid opcodes. When one of these invalid opcodes is decoded, an exception (often called Illegal Instruction exception) is triggered. If we can assume that the application code is correct and no fault will happen in the system, we can label faults associated to the logic in charge of triggering this exception as safe. In other words, we can identify a constraint, stating that any illegal opcode will never enter the CPU decoding logic.

**RESET LOGIC**. This category of safe faults relates to the *reset of the flip flops*. In all processors there is a module in charge of driving the reset signal entering all or most of the flip flops. This module activates the signal when several conditions become trues. For example, when an external asynchronous reset is activated, at power-on, or when a reset instruction is executed. In most system we can safely assume that no one of these conditions will happen during the normal system operations. Hence, most faults related to this logic can be labeled as safe.

**UNUSED INSTRUCTIONS**. Once the application code used by the system is known, an analysis can be done to check whether any instruction supported by the processor is possibly unused. An instruction never appears in the code may happen for different reasons, possibly connected also to non-functional requirements, including the following:

- The programmer decided to avoid its usage. As a common example, it may happen that the programmer decided not to use floating point instructions, although the processor supports them (and includes a floating point unit), either because floating point is not required by the application.
- The compiler didn't use the instruction, e.g., because it was not necessary. As a simple example, in some applications you cannot find any load or store instruction with byte parallelism (but only with word parallelism), since no 8-bit variable is used in the code. Division is another instruction which is sometimes not required, and hence not used.

The results reported in [5] clearly show that even with complex applications the percentage of unused hardware in a processor can be relevant.

| Unit | #Faults | Safe faults categories ||||||  Total safe faults removed ||
|---|---|---|---|---|---|---|---|---|---|
|  |  | Reset logic | SPR addressing | Memory access | PC update logic | Decoding logic | Unused instructions |  |  |
| CPU | 115,137 | 2,888 | 1,610 | 62 | 270 | 234 | 352 | 5,434 | 4.72% |
| alu | 10,967 | 0 | 0 | 0 | 0 | 0 | 0 | 0 | 0% |
| cfgr | 196 | 0 | 0 | 0 | 0 | 0 | 0 | 0 | 0% |
| ctrl | 3,998 | 112 | 0 | 0 | 0 | 192 | 18 | 322 | 8.05% |
| exception | 6,685 | 342 | 0 | 0 | 0 | 42 | 0 | 384 | 5.74% |
| freeze | 142 | 9 | 0 | 0 | 0 | 0 | 0 | 9 | 6.34% |
| genpc | 3,712 | 123 | 0 | 0 | 0 | 0 | 0 | 123 | 3.31% |
| if | 2,565 | 198 | 0 | 0 | 270 | 0 | 0 | 468 | 18.25% |
| lsu | 2,519 | 0 | 0 | 62 | 0 | 0 | 0 | 62 | 2.46% |
| mult_mac | 35,441 | 789 | 0 | 0 | 0 | 0 | 352 | 1,141 | 3.22% |
| operandmuxes | 3,120 | 82 | 0 | 0 | 0 | 0 | 0 | 82 | 2.63% |
| rf | 38,118 | 1,072 | 160 | 0 | 0 | 0 | 0 | 1,232 | 3.23% |
| sprs | 5,564 | 72 | 1,450 | 0 | 0 | 0 | 0 | 1,522 | 27.35% |
| wbmux | 2,070 | 38 | 0 | 0 | 0 | 0 | 0 | 38 | 1.83% |

*Table 1: Safe faults identification results*

## 4. Safe faults identification method

In this paper, a method is proposed to identify the safe faults belonging to the categories described in the previous section. Other safe faults can obviously exist, and other methods can be used for their identification.
The proposed method is based on the following steps:
1. Constraints extraction at the system level
2. Identification of the involved modules and translation of the system-level constraints into input and output constraints on the single module
3. ATPG activation.

Step 1 corresponds to the analysis performed in the previous Section. This step is performed manually. Clearly, the constraints that we can observe in a system are not limited to those we listed here, and further constrains can be identified, depending on the hardware and software characteristics of the system. Step 2 is also performed manually, and aims at translating the general constraints at the system level into simpler constraints affecting input or output signals at the module level. In other words, some of the considered constraints can be expressed stating that a given input or output signal of a given module always remains at a given value during the system operational life. Other constraints may imply that a given input or output configuration is never reached during the operational phase. Moving from system to module level is crucial to simplify the task of the ATPG used in the next step and make it computationally feasible. Step 3 leverages the ATPG which is asked to identify at the module level (i.e., considering the module as an isolated fully controllable and fully observable entity) the faults that become untestable due to the specified input and/or output constraints. Being based on the ATPG, this step is fully automatic. Depending on the constraint, two cases are consider:
1. Constraints affecting a single input or output signal of a module: in this case we can straightforwardly connect this signal to the fixed values 0 or 1, and then run the ATPG.
2. Constraints stating that a given combination will never appear on the input signals of the module (as in the case of an unused instruction). In this case, we can modify the netlist of the module by artificially adding some extra circuitry which is fed with the same input signals affected by the forbidden configuration and generates an extra output signal, which holds the value 1 iff the forbidden combination is applied. By forcing this additional output signal to 0 we can push the ATPG to identify untestable faults in the module, stemming from the considered constraint. The same approach can clearly be extended to the case in which multiple input combinations never appear on the inputs.

### 4.1. Experimental Results

In this section we report some experimental results computed using the method proposed in Section 4 on the safe faults categories described in Section 3. Table 1 shows the overall results that will be discussed in this chapter, in particular the number of safe faults identified for each CPU unit. The last subsection shows the impact of the identified safe faults on the achievable Fault Coverage. As a case study, an embedded system based on an OpenRISC 1200 [12] is was chosen. The implementation used by us is equipped with 2 MB of RAM for data and 2 MB of flash memory for code. The OR1200 we used has been synthetized with the NanGate 45nm library resulting in 115,137 possible stuck-at faults.

### 4.2. Safe faults identification

Concerning the first category (Special Purpose Registers), we assume that the application program does not generate IDs associated with registers not present in the processor and that the application program does not use the peripherals or I/O interfaces. With these assumptions, it is possible to impose a constraint on the most significant 5 output bits of the *spr_addr* bus: these 5 bits are fixed to the logical zero value. Launching a commercial ATPG tool with the above constraint on the *sprs* module we could identify 1,450 safe faults, as shown in Table 1. The same constraints imposed on the output signals of the *sprs* unit can be imposed on the input signals of the *Register File* unit. With this constraints, 160 safe faults in the register file unit are identified. Moving to the *data memory access* category, in the OR1200 the address bus has a 32-bit parallelism, able to index 4GB of RAM memory. In the implementation of the OR1200 used, the RAM memory addresses range from 0x00100000 to 0x002FFFFF. Hence, the 11 most significant bits of the RAM address bus are fixed to 0 for hardware constraints. With the method proposed applied to the load store unit (*lsu*) with the constraints found on RAM addressing, 62 safe faults are identified, as shown in Table 1. Similar considerations can be made for the *Program Counter update logic* category. The address bus of the flash memory has a 32-bit parallelism. In the memory map of the OR1200 system we considered, the flash memory is mapped to addresses ranging from 0x04000000 to 0x05FFFFFF. Hence, the 5 most significant bits of the flash address bus are fixed to 0. With the method proposed, on the instruction fetch unit (*if*) and with the constraints imposed on its output signals, 270 safe faults were identified, as shown in Table 1. The method proposed is applied to the *decoding of the instruction opcode* category. In this case, we identified an output signal of the decode unit (*ctrl* in the OR1200 model) which is activated when an Illegal Instruction exception is triggered. The identified constraint corresponds to forcing this signal to 0. By running the ATPG with this constraint on the *ctrl* unit we identified 192 faults as safe. The same signal triggered by the Illegal Instruction exception is directly connected to the interrupt controller. Hence, the same constraint can be used on the input of the exception unit (*exception*). Once again, the signal is forced to zero. With this constraint and the proposed method, 42 safe faults are identified in the exception unit. As described in Section 4, we can assume that during the normal operational phase the reset signal entering each flip flop is never activated. Hence, we can force this signal to 0 and identify the faults that become untestable as a consequence. By forcing the reset signal entering into the different units to 0 and then running the ATPG on each of them we could identify 2,888 safe faults. For the unused instruction category, we selected two sample instructions performing integer division (*l.div* and *ldivu*). The proposed method applied on the *mult_mac* and *ctrl* units, amounting to 352 and 18, respectively.

### 4.3. Fault Coverage without Safe Faults

Table 1 shows how many safe faults have been identified for each unit inside the processor and for each category of safe faults considered here. The column *Total safe faults removed* reports the total numbers of safe faults identified for each unit and their percent with respect to the number of faults of the unit. Overall, 5,434 safe faults have been identified in the OR1200 CPU. As a major result of this effort, it is now possible to compute the Fault Coverage achieved by a generic test step by considering a list of faults from which safe faults have been removed. For this purpose, we can use the *Fault Coverage* without Safe Faults (FC_safe) metric, as done already in [4], computed via the expression:

**FC_safe = #detected faults / (#faults - #safe faults)**

The proposed method for the identification of safe faults and for the calculation of the FC_safe is independent of the test method chosen to test the device. In our case study we chose to test the OR1200 CPU by resorting to a Self-Test Library (STL) based on the Software-Based Self-Test (SBST) paradigm. Details about the techniques followed to generate the adopted STL can be found in [4]. This STL consists of 13 test programs, providing an initial stuck-at fault coverage of 83.95%. Details on the number of faults labelled as detected and not detected are shown in Table 2. When computing the FC_safe we got a figure of 88.14%. Table 2 shows the details of the FC_safe for each unit of the CPU.

### 5. Conclusions

In this paper we focused on Safe Faults, and described first some categories of Safe Faults that can be found in embedded processors. We also proposed a method to partly automate their identification.

### 6. Acknowledgements

The work has been supported by project H2020 MSCA ITN RESCUE (EU Horizon 2020, Grant 722325), Estonian research grant IUT 19-1 and Excellence Centre EXCITE in Estonia, in collaboration with the Politecnico di Torino in Italy.

| Unit | Initial Fault Coverage | | | Final Fault Coverage | |
|---|---|---|---|---|---|
| | #Faults | FC [%] | #Detected faults | #Safe Faults | FC_safe [%] |
| CPU | 115,137 | 83.95 | 96,696 | 5,434 | 88.14 |
| alu | 10,967 | 94.98 | 10,417 | 0 | 94.98 |
| cfgr | 196 | 0 | 0 | 0 | 0 |
| ctrl | 3,998 | 83.38 | 3,346 | 322 | 91.02 |
| exception | 6,685 | 16.60 | 1,116 | 384 | 17.71 |
| freeze | 142 | 67.96 | 98 | 9 | 73.68 |
| genpc | 3,712 | 53.83 | 1,998 | 123 | 55.67 |
| if | 2,565 | 66.14 | 1,698 | 468 | 80.97 |
| lsu | 2,519 | 74.67 | 1,881 | 62 | 76.56 |
| mult_mac | 35,441 | 94.16 | 33,381 | 1,141 | 97.32 |
| operandmuxes | 3,120 | 96.12 | 2,999 | 82 | 98.72 |
| rf | 38,118 | 93.85 | 35,774 | 1,232 | 96.98 |
| sprs | 5,564 | 40.38 | 2,250 | 1,522 | 55.66 |
| wbmux | 2,070 | 82.03 | 1,698 | 38 | 83.56 |

*Table 2: Fault Coverage result*